\documentstyle[12pt]{article}   
\begin{document}
\begin{center}   
\Large
{\bf{Theory of non-stationary activated rate \\
processes : nonexponential
relaxation kinetics}}
\end{center}

\begin{center}
{\bf{Jyotipratim Ray Chaudhuri$^{\rm a}$, Gautam Gangopadhyay$^{\rm b}$,\\ 
Deb Shankar Ray$^{\rm a}$}}
\end{center}

\begin{center}
$^{\rm a}${\bf{Indian Association for the Cultivation of Science}}\\
{\bf{Jadavpur, Calcutta 700 032, INDIA.}}

$^{\rm b}${\bf{S. N. Bose National Centre for Basic Sciences}}\\
{\bf{JD Block, Sector III, Salt Lake City, Calcutta 700 091, INDIA.}}
\end{center}

\begin{abstract}
We have explored a simple microscopic model to simulate a thermally activated
rate process where the associated bath which comprises a set of relaxing 
modes is not in an equilibrium state. The model captures some of the essential 
features of non-Markovian Langevin dynamics with a fluctuating barrier. Making 
use of the Fokker-Planck description we calculate the barrier dynamics in the 
steady state and non-stationary regimes. The 
Kramers-Grote-Hynes reactive 
frequency has been computed in closed form in the steady state to 
illustrate the strong dependence of the dynamic coupling of the system with 
the relaxing modes. The influence of nonequilibrium excitation of the bath 
modes and its relaxation on the kinetics of activation of the system mode is 
demonstrated. We derive the dressed time-dependent Kramers rate in the 
nonstationary 
regime in closed analytical form which exhibits strong non-exponential 
relaxation kinetics of the reaction co-ordinate. The feature can be 
identified as a typical non-Markovian dynamical effect.
\end{abstract}

\newpage

\begin{center}
\bf{I.\hspace{0.2cm}Introduction}
\end{center}

\vspace{0.5cm}

More than half a century ago Kramers$^{1}$ considered the problem of activated 
rate processes by using a model Brownian particle trapped in a one dimensional 
well which is separated by a barrier of finite height from a deeper well.
The particle was supposed to be immersed in a medium such that the medium 
exerts a frictional force on the particle but at the same time thermally 
activate it so that the particle may gain enough energy to cross the barrier.
Over several decades the model has been the standard paradigm in many areas
of physics and chemistry$^{2}$. The Kramers problem was to find the rate of 
escape from the well to the barrier. The motion of the particle is governed 
by the following phenomenological Langevin equation,
\begin{equation}
\ddot{x}=-\frac{1}{m}\frac{\partial V(x)}{\partial x} - \gamma\dot{x}
+ \frac{1}{m} F(t) \hspace{0.2cm},
\end{equation}
\noindent
where $x$ is the coordinate of the particle of mass $m$ moving in a potential
$V(x)$. $\gamma$ and $F(t)$ are the damping rate and the Gaussian stationary 
random force provided by the thermal bath respectively. The properties of 
noise can be summarized by the following two relations,
\begin{equation}
\langle F(t)\rangle=0 \hspace{0.4cm}, \hspace{0.4cm}
\langle F(0)F(t)\rangle=2 \gamma mKT \delta(t) \hspace{0.2cm}.
\end{equation}

The Langevin equation (1) is equivalent to the Fokker-Planck equation for 
probability distribution $p=p(x,v,t)$ [also known as Kramers equation],
\begin{equation}
\frac{\partial p}{\partial t}=\frac{1}{m}\frac{\partial V(x)}{\partial x}
\frac{\partial p}{\partial v}-v\frac{\partial p}{\partial x} + \gamma
\left[\frac{KT}{m} \frac{\partial^{2} p}{\partial v^{2}} +\frac{\partial}
{\partial v}(vp) \right] \hspace{0.2cm}.
\end{equation}

Kramers$^{1}$ obtained the steady state escape rate $k$ in the limiting cases 
of high and low damping rates in the following form,
\begin{equation}
k=\left\{\begin{array}{lllll}
\frac{\omega_{0}\omega_{b}}{2\pi\gamma}\exp[-\frac{E_{b}}{KT}] &  &  & 
\gamma\longrightarrow\infty  \\
\gamma\frac{E_{b}}{KT}\exp[-\frac{E_{b}}{KT}] & & & \gamma\longrightarrow 0
\end{array}\right. \hspace{0.2cm},
\end{equation}
\noindent
where $\omega_{o}$ and $\omega_{b}$ are the frequencies associated with the
curvature of the potential at the bottom of the well and at the barrier top,
respectively. $E_{b}$ refers to the depth of the well. Kramers has also
derived an expression for `intermediate' value of $\gamma$ :
\begin{eqnarray*}
k=\frac{\omega_{0}}{2\pi\omega_{b}}\left\{\left[ \left(\frac{\gamma}{2}
\right)^{2}+\omega_{b}^{2}\right]^{\frac{1}{2}}-\frac{\gamma}{2}\right\}
\exp(-E_{b}/KT)\hspace{0.2cm}.
\end{eqnarray*}

For non-Markovian random processes where one takes into account of the short
internal time scales of the system compared to that of the thermal bath, the 
Langevin equation(1) gets replaced by its non-Markovian counterpart$^{3,4}$, 
sometimes called the generalized Langevin equation (GLE);
\begin{equation}
\ddot{x}=-\frac{1}{m}\frac{\partial V(x)}{\partial x}-\int_{0}^{t}d\tau Z(t-\tau)
\dot{x}(\tau) + \frac{1}{m}R(t) \hspace{0.2cm},
\end{equation}
\noindent
where $R(t)$ is Gaussian but non-Markovian such that
\begin{equation}
\langle R(t) \rangle = 0 ,\hspace{1.0cm}\langle R(0)R(t) \rangle = Z(t)mKT
\hspace{0.2cm}.
\end{equation}
\noindent
The memory function $Z(t)$ is expressed in terms of Fourier-Laplace components
\begin{equation}
Z_{n}(\omega) = \int_{o}^{\infty} dt Z(t) e^{-in\omega t}
\end{equation}
\noindent
with \hspace{6.0cm}$Z_{0}(\omega) = \gamma$

Based on equation (5) Adelman$^{5}$ obtained the generalized Fokker-Planck 
equation for a Brownian oscillator with a parabolic potential as given by ;
\begin{equation}
\frac{\partial p}{\partial t} = -{\bar{\omega}}_{b}^{2} x \frac{\partial p}
{\partial v} -v\frac{\partial p}{\partial x}
+ {\bar{\gamma}}\frac{\partial}{\partial v} (vp)+ {\bar{\gamma}}
\frac{KT}{m}\frac{\partial^{2}p}{\partial v^{2}} + \frac{KT}{m}\left(\frac
{{\bar{\omega}}_{b}^{2}}{\omega_{b}^{2}}-1\right)\frac{\partial^{2}p}
{\partial v\partial x} \hspace{0.2cm},
\end{equation}
\noindent
where ${\bar{\gamma}}$ = ${\bar{\gamma}}(t)$ and ${{\bar{\omega}}_{b}^{2}}
={{\bar{\omega}}_{b}^{2}}(t)$ are now functions of time [although bounded
, they may not always provide long time limits] which play a decisive role
in the calculation of non-Markovian Kramers rate.

Various workers have made use of generalized Langevin equation to treat
the different aspects of the escape problem in the non-Markovian regime.
For example, Grote and Hynes$^{4}$ considered the average motion of the 
particle in the vicinity of the barrier governed by GLE and found that 
on the average the particle is slowed down by friction and defining a 
reactive frequency $\lambda_{r}$ they showed that the average motion 
goes as $\exp(\pm\lambda_{r} t)$.
The analysis of H\"anggi and Mojtabai$^{6}$ on the other hand is 
based on the generalized Fokker-Planck equation of Adelman with a parabolic 
potential in the high friction limit. The generalized FP approach has also 
been adopted by Carmeli and Nitzan$^{7}$ to derive the expression for the 
steady-state escape rate in the high and low friction limit in the Markovian 
as well as non-Markovian regimes. A comprehensive overview has been given in 
Ref.(2).

While the early post-Kramers development as summarized above is largely phenomenological, an 
interesting advancement in the theory of activated rate processes was made
when the generalized Langevin equation was realized in terms of a microscopic
model which comprises a system coupled linearly to a discrete set of harmonic
oscillators. Using the properties of the bath and a normal mode analysis
it was shown$^{8}$ that the reactive frequency $\lambda_{r}$ defined by Grote 
and Hynes$^{4}$ for the average motion across the barrier is actually a 
renormalised effective barrier frequency.

The object of the present paper is twofold : First is
to consider a simple variant of the 
system-heat bath model$^{9,10,11}$ to simulate the activated rate processes, 
where the associated bath is in a nonequilibrium state. The model incorporates
some of the essential features of Langevin dynamics with a fluctuating
barrier which had been heuristically and phenomenologically proposed earlier
in several occasions.$^{10,13-17}$
While the majority of the treatments of the phenomenological
fluctuating barrier rest on the reduction of the equations to overdamped 
limit$^{5,10,14}$, thus restricting the validity of the solutions in the
large time limit, we take full account of the inertial terms in our calculation
of barrier dynamics and probability distribution function both in the long
time and in the short time nonstationary regimes. The Fokker-Planck
description allows us to calculate Kramers-Grote-Hynes reactive frequency
pertaining to these situations for non-Markovian dynamics in closed form. 
Second, since
the theories of activated processes traditionally deal with
stationary bath, the nonstationary activated processes has remained largely
overlooked so far. We specifically address this issue and examine the influence
of initial excitation and subsequent relaxation of a bath modes on the 
activation of the reaction co-ordinate. We show that relaxation of the nonequilibrium
bath modes may result in strong non-exponential kinetics and a nonstationary
Kramers rate. The physical situation that has been addressed is the following :

We consider that at $t=0_-$, the time just before the system and the bath is
subjected to an external excitation, the system is appropriately thermalized.
At $t=0$, the excitation is switched on and the bath is thrown into a 
nonstationary state which behaves as a nonequilibrium reservoir. We follow the 
stochastic dynamics of the system mode after $t>0$. The important separation
of the time scales of the fluctuations of the nonequilibrium bath and the
thermal bath (to which it relaxes) is that the former effectively remains
stationary on the fast correlation of the thermal noise.

The outline of the paper is as follows;
Following Ref. [10] we discuss in Sec.II a microscopic model to simulate an 
activated rate
process where the system in question is not initially thermalized. Appropriate
elimination of reservoir degrees of freedom leads to a nonlinear non-Markovian
Langevin equation which governs the dynamics of a particle with a fluctuating
barrier, stochasticity being contributed by both (additive) thermal noise
and a slower (multiplicative) noisy relaxing nonequilibrium modes. The 
Fokker-Planck description is provided in Sec.III. The standard Markovian 
description and the generalized FP equation of Adelman's form can be 
recovered in the appropriate limits. In Sec.IV we derive the expression for 
Kramers rate of barrier crossing in the non-Markovian but steady state regime and 
show that the Kramers-Grote-Hynes ``reactive frequency'' can be explicitly 
realized in this model in closed form. Sec.V is devoted to nonstationary 
aspect. We solve the time-dependent FP equation for nonstationary probability 
density and calculate the corresponding current. An expression for Kramers 
rate in the nonstationary regime in closed analytical form is derived. The 
paper is concluded in Sec.VI.

\vspace{0.5cm}

\begin{center}
\bf{II.\hspace{0.2cm} The model and the Langevin equation}
\end{center}

\vspace{0.5cm}

We consider a model consisting of a system mode coupled to a set of relaxing 
modes considered as a semi-infinite dimensional system 
(\{$q_k$\}-subsystem) which effectively 
constitutes a nonequilibrium bath. This, in turn, is in contact with a 
thermally equilibrated reservoir. Both the reservoirs are 
composed of two sets of harmonic oscillators characterized by the frequency 
sets $\{\omega_{k}\}$ and $\{\Omega_{j}\}$ for the nonequilibrium and the 
equilibrium bath, respectively. The system-reservoir combination evolves 
under the total Hamiltonian
\begin{eqnarray}
H=\frac{p^{2}}{2m}+V(x)+\frac{1}{2}\sum_{j}(P_{j}^{2}+\Omega_{j}^{2}Q_{j}^{2})
+\frac{1}{2}\sum_{k}(p_{k}^{2}+\omega_{k}^{2}q_{k}^{2})\nonumber\\
-x\sum_{j}K_{j}Q_{j}-g(x)\sum_{k}q_{k}-\sum_{j,k}\alpha_{jk}q_{k}Q_{j}
\hspace{0.2cm},
\end{eqnarray}
\noindent
the first two terms on the right hand side describe the system mode. The
Hamiltonian for the thermal and nonequilibrium baths are described by the 
sets $\{Q_{j},P_{j}\}$ and $\{q_{j},p_{j}\}$ for coordinates and momenta,
respectively. The coupling terms containing $K_{j}$ refers to the usual
system-thermal bath linear coupling. The last two terms indicate the coupling
of the nonequilibrium bath to the system and the thermal bath modes, 
respectively. Since in the present problem, H is considered to be classical 
and temperature, T high for the thermally activated problem we note that 
quantum effects do not play any significant role. Hamiltonian (9) 
is a simpler variant of that treated in Ref.[10]. For simplicity we take 
$m=1$ in (9) and for rest of the treatment. As shown in Ref. [10] the model 
(9) captures the essential features of fluctuating barrier dynamics. We
recall the relevant aspect in the following discussions.

Eliminating the equilibrium reservoir variables $\{Q_{j},P_{j}\}$ in an
appropriate way $^{9,10}$ one may show that the nonequilibrium bath modes 
obey the following equations of motion,
\begin{equation}
{\ddot{q}}_{k}+\gamma{\dot{q}}_{k}+\omega_{k}^{2}q_{k}=g(x)+\eta_{k}(t)
\hspace{0.2cm}.
\end{equation}

This takes into account of the average dissipation ($\gamma$) of the
nonequilibrium reservoir modes $q_{k}$ due to its coupling to thermal 
reservoir which induces fluctuations $\eta_{k}(t)$ characterized by
$\langle \eta_{k}(t)\rangle=0$ and the usual fluctuation-dissipation theorem
$\langle \eta_{k}(t)\eta_{k}(0)\rangle=2\gamma KT\delta(t)$. We mention here
that moving from Eq.(9) to (10) generate cross terms of the form
$\sum_j \gamma_{kj} q_j$, which are neglected for $j \neq k$.

Proceeding similarly to eliminate the thermal reservoir variables from the 
equations of motion of the system mode one obtains
\begin{equation}
{\ddot{x}}+\gamma_{\rm eq}{\dot{x}}+V'(x)=\xi_{\rm eq}(t)+g'(x)\sum_{k}q_{k}
\hspace{0.2cm},
\end{equation}
\noindent
where $\gamma_{\rm eq}$ refers to the dissipation coefficient of the system 
mode due to its direct coupling to the thermal bath providing fluctuations
$\xi_{\rm eq}(t)$. Here we have
\begin{eqnarray*}
\langle \xi_{\rm eq}(t)\rangle=0 \hspace{0.2cm}{\rm and} \hspace{0.2cm}
\langle \xi_{\rm eq}(t)\xi_{\rm eq}(0)\rangle=2\gamma_{\rm eq}KT\delta(t)
\hspace{0.2cm}.
\end{eqnarray*}

Now making use of the formal solutions of Eq.(10)$^{10}$ which takes into 
account of the relaxation of the nonequilibrium modes and integrating over 
the nonequilibrium modes with a Debye type frequency distribution of the form,
\begin{eqnarray*}
\rho(\omega)=\left\{ \begin{array}{lllll}
3\omega^{2}/2\omega_{c}^{3} & & , & {\rm for} |\omega| \le \omega_{c}\\
0 & & , & {\rm for}|\omega| > \omega_{c}
\end{array} \right.
\end{eqnarray*}
\noindent
where $\omega_{c}$ is the high frequency Debye cut-off, one finally arrives
at the following Langevin equation of motion for the system mode,
\begin{equation}
{\ddot{x}}+\Gamma(x){\dot{x}}+\tilde{V}'(x)=\xi_{\rm eq}(t)+g'(x)\xi_{\rm neq}
(t)\hspace{0.2cm}.
\end{equation}

Here $\Gamma(x)$ is a system coordinate dependent dissipation constant 
composed of $\gamma_{\rm eq}$ and $\gamma_{\rm neq}$ as follows,
\begin{equation}
\Gamma(x)=\gamma_{\rm eq}+\gamma_{\rm neq}[g'(x)]^{2}\hspace{0.2cm}.
\end{equation}
\noindent
$\xi_{\rm neq}$ refers to the fluctuations of the nonequilibrium bath modes 
which effectively cause a damping of the system mode by an amount 
$\gamma_{\rm neq}[g'(x)]^{2}$.

Eq.(12) also includes the modification of the potential $V(x)$ in which the
particle moves as
\begin{equation}
{\tilde{V}}(x)=V(x)-\frac{\omega_{c}}{\pi} \; \gamma_{\rm neq} \; g^{2}(x)
\hspace{0.2cm}.
\end{equation}

Eq.(12) thus describes the effective dynamics of a particle in a modified
barrier, where the metastability of the well originates from the dynamic
coupling $g(x)$ of the system mode with the nonequilibrium bath modes. 
It is necessary to stress here that $g(x)$, in general, is nonlinear. This
nonlinearity has two immediate consequences. First, by virtue of the term
${\tilde{V}}'(x)$ in Eq.(12) it gives rise to a fluctuating barrier. Second,
the term $g'(x) \xi_{\rm neq} (t)$ imparts a multiplicative noise term 
in Eq.(12)
in addition to the usual additive noise term $\xi_{\rm eq} (t)$. 
We point out here that the problem
of diffusion over a fluctuating barrier$^{13-17}$ of similar nature 
has been addressed earlier by a 
number of workers from the phenomenological point of view. For example, Stein 
et.al.$^{14}$ have calculated the decay of probability from the metastable 
state in the white noise limit and also for short finite correlation times 
for the fluctuating part of the potential. Riemann and Elston$^{15}$ have 
calculated an asymptotic rate formula when the particle is subjected to both 
dichotomous and thermal noise.

The treatment followed in the aforesaid cases concerns overdamped situation
and, in general, the validity is restricted to long time limit. In the present 
problem, however, we look at the stochastic process right from the moment
the nonequilibrium excitation (followed by the relaxation) sets in. We are
therefore forced to take into consideration of the inertial term in Eq.(12)
on its usual footing.

We now turn to the another aspect of the problem.
In order to define the problem described by Eq.(12) completely, it is further
necessary to state the properties of fluctuations of the nonequilibrium bath
$\xi_{\rm neq}(t)$. We have first for Gaussian noise
\begin{eqnarray*}
\langle \xi_{\rm neq} (t) \rangle = 0 \; \; .
\end{eqnarray*}
\noindent
Also the essential properties of $\xi_{\rm neq} (t)$ explicitly depend on the
nonequilibrium state of the intermediate oscillator modes $\{ q_k \}$
through ${\cal U}(\omega, t)$, the energy density distribution function at time
$t$ in terms of the following fluctuation-dissipation relation$^{10}$ for the
nonequilibrium bath,
\begin{eqnarray}
{\cal U}(\omega , t) & = & \frac{1}{4\gamma_{\rm neq}} \int_{-\infty}^{+\infty}
d\tau \; \langle \xi_{\rm neq} (t) \xi_{\rm neq} (t+\tau) \rangle \;
e^{i\omega \tau} \nonumber \\
& = & \frac{1}{2} KT \; + \; e^{-\gamma t/2} \; 
\left [ {\cal U}(\omega,0) - \frac{1}{2} KT \right ] \; \; ,
\end{eqnarray}
\noindent
$\left [ {\cal U}(\omega,0) - \frac{1}{2} KT \right ] $ is a measure of departure of
energy density from thermal average at $t=0$. The exponential term implies
that this deviation due to the initial excitation decays asymptotically to
zero as $t\rightarrow \infty$, so that one recovers the usual 
fluctuation-dissipation relation for the thermal bath. 
With the above specification of correlation function of $\xi_{\rm neq}$
Eq.(15) thus attributes
the nonstationary character of the \{$q_k$\}-subsystem.

In passing, we stress that the above derivation$^{10}$ is based on the assumption
that $\xi_{\rm neq}$ is effectively stationary on the fast correlation of the 
thermal modes. This is a
necessary requirement for the systematic separation of time scales 
involved in the dynamics. We point out that the effective dynamics sets no
choice on any special form of coupling $g(x)$ between the system mode and the 
relaxing mode and as such this may be of arbitrary nonsingular type for our
problem we have considered here.

\newpage

\begin{center}
\bf{III.\hspace{0.2cm}The generalized Fokker-Planck description}
\end{center}

\vspace{0.5cm}

Eq.(12) is the required Langevin equation for the particle moving in a
modified potential ${\tilde{V}}(x)$ [Eq.(14)] and 
damped by a coordinate-dependent
friction $ \Gamma (x) $ [Eq.(13)] due to its linear coupling to a thermal
bath and nonlinear coupling to the $\{q_k\}$-subsystem characterized by
fluctuations $\xi_{\rm neq} (t)$. Before proceeding further a few pertinent
points are to be noted to stress some distinct and important aspects
of the model.

First, depending on the system-\{$q_k$\}-subsystem coupling $g(x)$ both the modified
potential ${\tilde{V}}(x)$ as well as $\Gamma(x)$ are, in general, nonlinear.
So the stochastic differential equation (12) is nonlinear. Again, the 
stochasticity in Eq.(12) is composed of two parts : $\xi_{\rm eq} (t)$ is an
additive noise due to thermal bath while $\xi_{\rm neq} g'(x)$ is a 
multiplicative contribution due to nonlinear coupling to \{$q_k$\}-subsystem.
It is thus important to note that the presence of multiplicative noise and
a fluctuating barrier are associated with nonlinearity in $g(x)$.

Second, the Langevin equation (12) is non-Markovian. The origin of this
non-Markovian nature lies in the decaying term in Eq.(15) where the decay
explicitly expresses the initial nonequilibrium nature of the 
$\{ q_k\}$-subsystem following the sudden excitation at $t=0$. This 
non-Markovian feature is thus not to be confused with that arises due to
the usual frequency dependence of the dissipation constant.

Third, although the modification of $V(x)$ is due to the specific choice of
the Debye model for the mode density which has so far been commonly used, the 
theory remains effectively unchanged as one goes over to more
complicated spectrum.

We now rewrite Eq.(12) in the form,
\begin{equation}
\left.\begin{array}{l}
\dot{u}_{1}=F_{1}(u_{1},u_{2},t ; \xi_{\rm neq},\xi_{\rm eq})
\dot{u}_{2}=F_{2}(u_{1},u_{2},t ; \xi_{\rm neq},\xi_{\rm eq})
\end{array}\right\}\hspace{0.2cm},
\end{equation}
\noindent
where we use the following abbreviations,
\begin{equation}
\left.\begin{array}{l}
u_{1}=x\\
u_{2}=v \end{array}\right\} 
\end{equation}
\noindent
and
\begin{equation}
\left.\begin{array}{l}
F_{1}=v\\
F_{2}=-\Gamma(x)v-\tilde{V}^{'}(x)+ \xi_{\rm eq}(t)+g'(x)
\xi_{\rm neq}(t)\end{array}\right\} \hspace{0.2cm}.
\end{equation}

\noindent
The vector $u$ with components $u_{1}$ and $u_{2}$ thus represents a point in
a 2-dimensional `phase space' and the Eq.(16) determines the velocity at each
point in this phase space. The conservation of points now asserts the 
following linear equation of motion for density $\rho(u,t)$ in `phase space',
\begin{eqnarray*}
\frac{\partial}{\partial t}\rho(u,t)=-\sum_{n=1}^{2}\frac{\partial}{\partial 
u_{n}} F_{n}(u,t;\xi_{\rm neq},\xi_{\rm eq})\rho(u,t)\hspace{0.2cm},
\end{eqnarray*}
\noindent
or more compactly
\begin{equation}
\frac{\partial \rho}{\partial t}=-\nabla\cdot F\rho\hspace{0.2cm}.
\end{equation}

Our next task is to find out a differential equation whose average solution 
is given by $\langle \rho \rangle$ where the stochastic averaging has to be 
performed over two noise processes $\xi_{\rm neq}$ and $\xi_{\rm eq}$. To
this end we note that $\nabla \cdot F$ can be partitioned into two parts ;
a constant part $\nabla \cdot F_0$ and a fluctuating part $\nabla \cdot F_1 (t)$,
containing these noises. Thus we write
\begin{equation}
\nabla \cdot F(u,t;\xi_{\rm neq},\xi_{\rm eq}) = \nabla\cdot F_0(u) +
\epsilon \nabla \cdot F_1 (u,t;\xi_{\rm neq},\xi_{\rm eq}) \; \; ,
\end{equation}
\noindent
where $\epsilon$ is a parameter (we put it as an external parameter to keep 
track of the order of the perturbation expansion in $\epsilon \tau_c$, where
$\tau_c$ is the correlation time of fluctuation of $\xi_{\rm neq} (t)$ ; we
put $\epsilon=1$ at the end of calculation) and also note that
$\langle F_1(t) \rangle =0$. Eq.(19) therefore takes the following form ,
\begin{equation}
\dot{\rho} (u,t) = (A_0 \; + \; \epsilon A_1) \; \rho (u,t) \; \; ,
\end{equation}
\noindent
where $A_0=-\nabla \cdot F_0$, $ A_1=-\nabla \cdot F_1$. The symbol $\nabla$
is used for the operator that differentiate everything that comes after it
with respect to $u$.

Making use of one of the main results for the theory of linear equation of the
form (21) with multiplicative noise, we derive an average equation for $\rho$
[$\langle \rho \rangle = p(u,t)$, the probability density of $u(t)$ ; for
details refer to Van Kampen$^{12}$],
\begin{equation}
\dot{p} =\left \{ A_0 + \epsilon^2 \int_0^\infty \langle A_1(t) \;
\exp (\tau A_0) \; A_1 (t-\tau) \rangle \; \exp (-\tau A_0) \right \} \;
p \; \; .
\end{equation}

The above result is based on second order cumulant expansion and is valid
in the case that fluctuations are small but rapid and the correlation time 
$\tau_c$ is short but finite, i.e., 
\begin{eqnarray*}
\langle A_1(t) \; A_1(t') \rangle =0 \; \; {\rm for} \; 
|t-t'| > \tau_c \; \; .
\end{eqnarray*}

\noindent
The Eq.(22) is exact in the limit correlation time $\tau_c$ tends to zero.
Using the expressions for $A_0$ and $A_1$ we obtain
\begin{eqnarray}
\frac{\partial p(u,t)}{\partial t} = \{ -\nabla\cdot F_0 \; + \;
\epsilon^2 \int_0^\infty  d\tau \; \langle \nabla\cdot F_1(t) \;
\exp(-\tau \nabla\cdot F_0) \; \nabla\cdot F_1(t-\tau) \rangle 
\nonumber \\
\exp(\tau\nabla\cdot F_0) \} \; p(u,t) \; \; .
\end{eqnarray}

\noindent
The operator $\exp(-\tau\nabla\cdot F_0)$ in the above equation provides the
solution of the equation 
\begin{equation}
\frac{\partial f(u,t)}{\partial t} = - \nabla \cdot F_0 \; f(u,t) \; \; ,
\end{equation}

\noindent
($f$ signifies the unperturbed part of $\rho$) which can be found explicitly
in terms of characteristics curves. The equation
\begin{equation}
\dot{u} = F_0(u)
\end{equation}

\noindent
for fixed $t$ determines a mapping from $u(\tau=0)$ to $u(\tau)$, i.e., 
$u\rightarrow u^\tau$ with inverse $(u^\tau)^{-\tau}=u$. The solution of 
Eq.(24) is 
\begin{equation}
f(u,t) = f(u^{-t},0) \left | \frac{d (u^{-t})}{d(u)} \right | =
\exp(-t \nabla\cdot F_0) f(u,0) \; \; ,
\end{equation}

\noindent
$\left | \frac{d (u^{-t})}{d(u)} \right |$ being a Jacobian determinant. The 
effect  of $\exp(-t\nabla\cdot F_0)$ on $f(u)$ is as follows ;
\begin{equation}
\exp(-t\nabla\cdot F_0) \; f(u,0) = f(u^{-t},0) 
\left | \frac{d (u^{-t})}{d(u)} \right | \; \; .
\end{equation}
\noindent
The above simplification when put in Eq.(23) yields
\begin{eqnarray}
\frac{\partial}{\partial t}p(u,t)=\nabla\cdot \left\{-F_{0}+\epsilon^{2}
\int_{0}^{\infty}\left|\frac{d(u^{-\tau})}{d(u)}\right|
\langle F_{1}(u,t)\nabla_{-\tau}\cdot F_{1}(u^{-\tau},t-\tau)
\rangle\right.\nonumber\\
\left.\left|\frac{d(u)}{d(u^{-\tau})}\right| d\tau\right\} p(u,t)\hspace{0.2cm}.
\end{eqnarray}
\noindent
$\nabla_{-\tau}$ denotes differentiation with respect to $u_{-\tau}$. We put
$\epsilon = 1$ for the rest of the treatment. We now identify,
\begin{equation}
\left.\begin{array}{l}
u_{1}=x\\
u_{2}=v\\
F_{01}=v\hspace{0.2cm},\hspace{0.2cm}F_{11}=0\\
F_{02}=-\Gamma(x)v-\tilde{V}'(x)\hspace{0.2cm},\hspace{0.2cm}
F_{12}=\xi_{\rm eq}(t)+g'(x)\xi_{\rm neq}(t) 
\end{array} \right \} \; \; .
\end{equation}
\noindent
In this notation Eq.(28) now reduces to
\begin{eqnarray}
\frac{\partial p}{\partial t}=-\frac{\partial}{\partial x}(vp)+\frac{\partial}
{\partial v}\left\{\Gamma v+\tilde{V}'(x)\right\}p\hspace{7.5cm}\nonumber\\
\nonumber\\
+\frac{\partial}{\partial v}\int_{0}^{\infty}d\tau\langle\left [\xi_{\rm eq}(t)
+g'(x)\xi_{\rm neq}(t)\right ]\left [\frac{\partial}{\partial v^{-\tau}}\{
\xi_{\rm eq}(t-\tau)+g'(x^{-\tau})\xi_{\rm neq}(t-\tau)\}\right]\rangle
p \hspace{0.2cm},
\end{eqnarray}
\noindent
where we have used the fact that the Jacobian obey the equation $^{12}$
\begin{equation}
\frac{d}{dt}\log\left|\frac{d(x^{t},v^{t})}{d(x,v)}\right|
=\frac{\partial}{\partial x}v+\frac{\partial}{\partial v}\{-\Gamma v+\tilde{V}
'(x)\} =-\Gamma 
\end{equation}
\noindent
so that Jacobian equals to $e^{-\Gamma t}$. 

As a next approximation we consider the `unpurterbed' part of Eq.(16) and 
take the variation of $v$ during $\tau_{c}$
into account to first order in $\tau_{c}$. Thus we have
\begin{equation}
x^{-\tau}=x-\tau v\hspace{0.2cm};\hspace{0.2cm}v^{-\tau}=v+\Gamma\tau v+\tau
\tilde{V}'(x)\hspace{0.2cm}.
\end{equation}

Neglecting terms ${\cal O}(\tau^{2})$ Eq.(32) yields,
\begin{equation}
\frac{\partial}{\partial v^{-\tau}}=(1-\Gamma\tau)\frac{\partial}{\partial v}
+\tau \frac{\partial}{\partial x}\hspace{0.2cm}.
\end{equation}

Taking into consideration of Eq.(33), Eq.(30) can be simplified after some 
algebra to the following form,
\begin{eqnarray}
\frac{\partial}{\partial t}p(x,v,t)=-\frac{\partial}{\partial x}(vp)+
\frac{\partial}{\partial v}\left\{ \Gamma (x)v+\tilde{V}'(x)-2g'(x)g''(x)
I_{nn}\right \}p\nonumber\\
\nonumber\\
+\left \{ I_{ee}+[g'(x)]^{2}I_{nn} \right \}\frac{\partial^{2} p}{\partial 
v\partial x}\hspace{5.0cm}\nonumber\\
\nonumber\\
+\left \{ J_{ee}-\Gamma(x)I_{ee}+[g'(x)]^{2}J_{nn}-\Gamma(x)[g'(x)]^{2}I_{nn}
-vg'(x)g''(x)I_{nn}\right\}\frac{\partial^{2} p}{\partial v^{2}}\hspace{0.2cm},
\end{eqnarray}
\noindent
where,
\begin{equation}
\left.\begin{array}{l}
I_{ee}=\int_{0}^{\infty} d\tau \langle\xi_{\rm eq}(t)\xi_{\rm eq}(t-\tau)\rangle\tau \\ 
I_{nn}=\int_{0}^{\infty} d\tau \langle\xi_{\rm neq}(t)\xi_{\rm neq}(t-\tau)\rangle\tau \\ 
J_{ee}=\int_{0}^{\infty} d\tau \langle\xi_{\rm eq}(t)\xi_{\rm eq}(t-\tau)\rangle \\ 
J_{nn}=\int_{0}^{\infty} d\tau \langle\xi_{\rm neq}(t)\xi_{\rm neq}(t-\tau)\rangle \\ 
\end{array}\right\}\hspace{0.2cm}.
\end{equation}

The subscripts $ee$ and $nn$ in the above expressions for the integrals over
the correlation functions refer to equilibrium and nonequilibrium baths, 
respectively. In deriving the last Eq.(34) we have assumed that the two 
reservoirs are uncorrelated. Eq.(34) is the required generalized Fokker-
Planck equation for our problem.

In order to allow ourselves a fair comparison with Fokker-Planck equation
of other forms$^{5,6,7}$, we first turn to the diffusion terms in Eq.(34). 
The coefficients are  coordinate $(x)$ dependent. It is customary to get rid of
this dependence by approximating the coefficients at the barrier top (say, $
x=0$) [one may also use mean field or steady state solutions of Eq.(34)
obtained by neglecting the fluctuation terms and putting appropriate
stationary condition in the diffusion coefficients].

The drift term in Eq.(34) refers to the presence of a dressed potential of the
form,
\begin{eqnarray*}
R(x)=\tilde{V}(x)-[g'(x)]^{2}\; I_{nn}
\end{eqnarray*}
\noindent
or
\begin{equation}
R(x)=V(x)-\frac{\omega_{c}}{\pi}
\gamma_{\rm neq} g^{2}(x)-[g'(x)]^{2} \; I_{nn} \; \; .
\end{equation}

The modification of the potential is essentially due to the nonlinear coupling
of the system to the nonequilibrium modes. $I_{nn}$ is a 
non-Markovian small contribution 
and therefore the third term in (36) may be neglected without any loss of
generality. For the rest of the treatment we use
$R(x)\simeq \tilde{V}(x)$. At the vicinity of the barrier top $x=0, \tilde{V}
'(x)$ may be approximated, as usual, by a parabolic potential, i.e.,
\begin{equation}
\tilde{V}(x)\simeq \bar{E}_{b}-\frac{1}{2}\bar{\omega}_{b}^{2}x^{2}
\end{equation}
\noindent
with
\begin{equation}
V(x)\simeq E_{b}-\frac{1}{2}\omega_{b}^{2}x^{2}\hspace{0.2cm}.
\end{equation}

\noindent
For convenience, one may set $g(0) = 0$ in the Taylor series expansion for
$g(x)$ (carried out at the barrier top $x=0$ ), without any loss of
generality. And one obtains 
\begin{equation}
\bar{E_{b}} = E_{b}
\end{equation}
and
\begin{equation}
\bar{\omega_b}^{2} = {\omega_b}^{2} + \frac{2 \omega_{c} \gamma_{neq}}{\pi} 
[g'(0)]^{2}\; \; .
\end{equation}

In the linearized description, the Fokker-Planck Eq.(34) is now reduced to
the following form,
\begin{equation}
\frac{\partial p}{\partial t}=-v\frac{\partial p}{\partial x}+\Gamma p+
[\Gamma v-\bar{\omega}_{b}^{2}x]\frac{\partial p}{\partial v}+
A\frac{\partial^{2} p}{\partial v^{2}}+B\frac{\partial^{2} p}{\partial v
\partial x}\hspace{0.2cm},
\end{equation}
\noindent
where we have used the following abbreviations;
\begin{equation}
A=J_{ee}-\Gamma(0)I_{ee}+[g'(0)]^{2}J_{nn}-\Gamma(0)[g'(0)]^{2}I_{nn}
\end{equation}
\noindent
and
\begin{equation}
B=I_{ee}+[g'(0)]^{2}I_{nn}\hspace{0.2cm}.
\end{equation}

From the last two relations we have
\begin{equation}
A=\left[ J_{ee}+g'(0)^{2}J_{nn}\right]-\Gamma(0)B
\end{equation}

Defining $A$ and $B$ as
\begin{equation}
A=\bar{\gamma}KT \hspace{0.2cm}{\rm and}\hspace{0.2cm}B=\bar{\beta}KT
\end{equation}
\noindent
one obtains
\begin{eqnarray}
\frac{\partial p}{\partial t}=-v\frac{\partial p}{\partial x}-\bar{\omega}_{b}
^{2}x\frac{\partial p}{\partial v}+\Gamma\frac{\partial}{\partial v}(vp)+
\bar{\gamma}KT\frac{\partial^{2} p}{\partial v^{2}}\nonumber\\
\nonumber\\
+KT\left[ \frac{J_{ee}+g'(0)^{2}J_{nn}}{\Gamma(0) KT}-\frac{\bar{\gamma}}
{\Gamma(0)}\right]\frac{\partial^{2} p}{\partial x\partial v}\hspace{0.2cm}.
\end{eqnarray}

Identifying
\begin{equation}
\bar{\Omega}^{2}=\Omega^{2}\left[ \frac{J_{ee}+g'(0)^{2}J_{nn}}{\Gamma(0) KT}
\right]\hspace{0.2cm},
\end{equation}
\noindent
Eq.(46) may be rewritten as,
\begin{eqnarray}
\frac{\partial}{\partial t}p(x,v,t)=-v\frac{\partial p}{\partial x}
-\bar{\omega}_{b}^{2}x\frac{\partial p}{\partial v}
+\Gamma\frac{\partial}{\partial v}(vp)+
\bar{\gamma}KT\frac{\partial^{2} p}{\partial v^{2}}\nonumber\\
\nonumber\\
+KT\left[ \frac{\bar{\Omega}^{2}(t)}{\Omega^{2}}-\frac{\bar{\gamma}}
{\Gamma(0)} \right ] \frac{\partial^{2} p}{\partial x\partial v}\hspace{0.2cm}.
\end{eqnarray}

\noindent
Here $\bar{\gamma}(t)$ and $\bar{\Omega}(t)$ are functions of time (due to
the relaxation of the nonequilibrium modes) as defined
by Eqs.(45) and (47). Or in other words 
nonstationary nature of the bath makes $\bar{\Omega}(t)$ time-dependent
through $J_{nn}$ term which is essentially a non-Markovian modification. 

Now the fluctuation-dissipation relations for equilibrium and nonequilibrium
baths stated in Sec.II may be invoked. For equilibrium baths as noted earlier 
we have the usual result; 
\begin{equation}
J_{ee}=\int_{0}^{\infty} d\tau \langle\xi_{\rm eq}(t)\xi_{\rm eq}(t-\tau)
\rangle=\gamma_{\rm eq}KT 
\end{equation}

For the nonequilibrium version, Eq.(15) may be rearranged further to
note that
\begin{equation}
J_{nn}=\int_{0}^{\infty} d\tau \langle\xi_{\rm neq}(t)\xi_{\rm neq}(t-\tau)
\rangle=\gamma_{\rm neq}KT(1+r e^{-\frac{\gamma}{2} t})
\end{equation}
\noindent
where $r$ is a measure of the deviation from equilibrium at the initial 
instant and is given by 
$r=\left \{ \frac{{\cal U}(\omega_{\rightarrow 0},0)}{2KT} -1 \right\}$.
Here ${\cal U}(\omega,t)$ defines the energy density distribution at time $t$.

Using (49) and (50) we obtain from Eq.(47)
\begin{equation}
\frac{\bar{\Omega}^{2}(t)}{\Omega^{2}}=1+\frac{r \gamma_{\rm neq}e^{-\frac{\gamma}{2} t}}
{\gamma_{\rm eq}+\gamma_{\rm neq}[g'(0)]^{2}}\hspace{0.2cm}.
\end{equation}

In the long time limit the relation reduces to
\begin{equation}
\left.{\cal L}t\right._{t \rightarrow \infty} \bar{\Omega}(t)=\Omega\hspace{0.2cm}.
\end{equation}

It is interesting to note that with the replacement $\frac{\bar{\Omega}^{2}
(t)}{\Omega^{2}} \sim \frac{\bar{\omega}_{b}^{2}}{\omega_{b}^{2}}$ (terms are 
of order $1+{\cal O}(\gamma)$) and $\Gamma(0) \sim \bar{\gamma}$ one recovers 
the Fokker-Planck equation in the Adelman's form$^{5}$ (Eq.(8)).

\vspace{0.5cm}

\begin{center}
\bf{IV.\hspace{0.2cm}Non-Markovian steady state Kramers rate}
\end{center}

\vspace{0.5cm}

We now proceed to analyze our generalized Fokker-Planck equation (48) and
calculate the steady state current and the Kramers escape rate over the
barrier. The procedure we follow in this section is similar to that of 
Kramers supplemented by H\"anggi and Mojtabai's earlier analysis$^{6}$.

As usual we make the ansatz
\begin{equation}
p(x,v,t)=F(x,v,t)\exp\left [ -\frac{\frac{v^{2}}{2}+\tilde{V}(x)}{KT}\right]
\end{equation}
\noindent
with $\tilde{V}(x)$ as approximated by a parabolic potential of the form
[ see Eqs. (37-40) ]
\begin{eqnarray*}
\tilde{V}(x)\simeq \bar{E}_{b}-\frac{1}{2}\bar{\omega}_{b}^{2}x^{2}
\end{eqnarray*}
\noindent
with$\hspace{5.0cm}\bar{E}_{b}=E_{b}$

\noindent
and$\hspace{5.0cm}\bar{\omega}_{b}^{2}=\omega_{b}^{2}+
\frac{2 \; \omega_{c} \; \gamma_{\rm neq}}{\pi} [g'(0)]^{2}$

\noindent
as stated earlier.

We seek an equation for $F$ of the form
\begin{equation}
F(x,v,t)=F(u,t)\hspace{0.2cm},\hspace{0.2cm}u=v+ax\hspace{0.2cm}.
\end{equation}

Inserting (53) and (54) in Eq.(48) we obtain
\begin{eqnarray}
\frac{\partial F}{\partial t}=\left\{(\Gamma-\bar{\gamma})-\frac{1}{KT}
(\Gamma-\bar{\gamma})v^{2}-\frac{\bar{\omega}_{b}^{2}}{KT}\Delta xv\right\}F
\nonumber\\
\nonumber\\
+\left[\left\{(\Gamma-2\bar{\gamma})-a(1+\Delta)\right\}v-\bar{\omega}_{b}^{2}
(1-\Delta)x\right]\frac{\partial F}{\partial u}\nonumber\\
\nonumber\\
+KT(\bar{\gamma}+\Delta a)\frac{\partial^{2}F}{\partial u^{2}}\hspace{0.2cm},
\end{eqnarray}
\noindent
where
\begin{equation}
\Delta=\frac{\bar{\Omega}^{2}(t)}{\Omega^{2}}-\frac{\bar{\gamma}}{\Gamma(0
)}\hspace{0.2cm}.
\end{equation}

Using (51), $\Delta$ may be rewritten as
\begin{equation}
\Delta\simeq \frac{r \gamma_{\rm neq}e^{-\frac{\gamma}{2} t}}{\gamma_{\rm eq}+\gamma_{
\rm neq}[g'(0)]^{2}}
\end{equation}
\noindent
for $\Gamma\sim\bar{\gamma}$.

Assuming $\frac{\Delta}{KT}$ and $(\Gamma-\bar{\gamma})$ to be very small
we obtain
\begin{equation}
\frac{\partial F}{\partial t} = KT\frac{\partial^{2} F}{\partial u^{2}} 
-\left[ \frac{\Gamma+a(1+\Delta)}{\Gamma+\Delta a}v+\frac{\bar{\omega}_{b}
^{2}(1-\Delta)}{\Gamma+\Delta a}x\right]\frac{\partial F}{\partial u} 
\hspace{0.2cm},
\end{equation}
\noindent
which may be written in the form
\begin{equation}
\frac{\partial F}{\partial t}=KT\frac{\partial^{2} F}{\partial u^{2}}+
\bar{\alpha}u\frac{\partial F}{\partial u}\hspace{0.2cm},
\end{equation}
\noindent
with
\begin{equation}
\bar{\alpha}=-\frac{\Gamma+a(1+\Delta)}{\Gamma+
\Delta a}\hspace{0.2cm},
\end{equation}

\noindent
and $a$ is a solution of the quadratic equation
\begin{equation}
a^{2}(1+\Delta)+\Gamma a-\bar{\omega}_{b}^{2}(1-\Delta)=0\hspace{0.2cm}.
\end{equation}

Since $\left.{\cal L}t\right._{t\rightarrow\infty}\Delta=0$, the long 
time or steady state solution of Eq.(58) is satisfied by
\begin{equation}
KT\frac{\partial^{2} F}{\partial u^{2}}+
\bar{\alpha}u\frac{\partial F}{\partial u}=0
\end{equation}
\noindent
with
\begin{equation}
\left.{\cal L}t\right._{t\rightarrow\infty}\bar{\alpha}(t)=-\frac{\Gamma+a}
{\Gamma}=\alpha({\rm say})\hspace{0.2cm}.
\end{equation}

Since the Eqs.(62) and (63) are identical in form to the expressions obtained
in the usual Kramers theory one can have the usual expressions for the
probability density $p(x,v)$ and the current $j_{s}$ as
\begin{equation}
p(x,v,\infty)=
N\left[\left(\frac{\pi KT}{2 \alpha}\right)^{\frac{1}{2}}
+\int_{0}^{v-|a|x}dz \exp\left(-\frac{\alpha z^{2}}{2KT}\right)\right]
\exp\left[ -\frac{\frac{v^{2}}{2}+\tilde{V}(x)}{KT}\right]
\end{equation}
\noindent
with
\begin{eqnarray*}
F_{s} = 
N\left[\left(\frac{\pi KT}{2 \alpha}\right)^{\frac{1}{2}}
+\int_{0}^{v-|a|x}dz \exp\left(-\frac{\alpha z^{2}}{2KT}\right)\right] \; \; ,
\end{eqnarray*}
\noindent 
(here the subscript $s$ in $F_s$ refers to steady state $F$) and
\begin{equation}
j_{s}=\int_{-\infty}^{+\infty}dv \hspace{0.1cm}vp(x,v)
=N(KT)^{\frac{3}{2}}\left(\frac{2\pi}{\alpha+1}\right)^{\frac{1}{2}}
\exp\left(-\frac{E_b}{KT}\right)\hspace{0.2cm},
\end{equation}
\noindent
where we have used the linearized version of $\tilde{V}(x)$ near the top of
the barrier at $x=0$,
\begin{eqnarray*}
\tilde{V}(x)=\bar{E}_{b}-\frac{1}{2}\bar{\omega}_{b}^{2}x^{2}\hspace{0.2cm},
\end{eqnarray*}
\noindent
with ${\bar{E}}_b = E_b$ and ${\bar{\omega}}_b$ is as given in Eq.(40)
and $N$ is the normalization constant.

Employing the asymptotic distribution (just before the system is subjected
to the shock at $t=0$) of $P_{w}(x,v)$ for 
$x\rightarrow -\infty$ and at $t=0_-$ from $p(x,v,t)$, where
$P_{w}(x,v)=p(x\rightarrow -\infty,v;t=0)$ [see Sec. V 
for calculation of $p(x,v,t)$], one obtains the total number of particles
in the well,
\begin{equation}
n_{a}=N \int_{-\infty}^{+\infty}dv \int_{-\infty}^{+\infty}dx P_{w}(x,v)
=N \frac{2\pi KT}{\omega_{0}}\left(\frac{2\pi KT}{\alpha}\right)^{\frac{1}{2}}
\hspace{0.2cm}.
\end{equation}
\noindent
Here $\omega_{0}$ is the frequency at the bottom of the left well. 
We have set
the potential energy at the bottom of the left well equal to zero, for
convenience.

The final result for the rate of escape in the steady state is given by
\begin{equation}
k=\frac{j_{s}}{n_{a}}=\frac{\omega_{0}\lambda}{2\pi\bar{\omega}_{b}}
e^{-E_{b}/KT}\hspace{0.2cm},
\end{equation}
\noindent
where
\begin{equation}
\lambda=\left[\left\{\left(\frac{\Gamma}{2}\right)^{2}+\bar{\omega}_{b}
^{2}\right\}^{\frac{1}{2}}-\frac{\Gamma}{2}\right]\hspace{0.2cm}.
\end{equation}

\noindent
It is evident that $\lambda$ is reminiscent of the
`reactive frequency' $\lambda_{r}$ of Grote and Hynes$^{4}$ . Microscopically
the non-Markovian character of the dynamics in $\lambda$ enters through the
explicit structure of $\Gamma$ and $\bar{\omega}_{b}$ which are given by
\begin{equation}
\Gamma=\gamma_{\rm eq}+\gamma_{\rm neq}[g'(0)]^{2}
\end{equation}
\noindent
and
\begin{equation}
\bar{\omega}_{b}^{2}=\omega_{b}^{2} \; + \; \frac{2\omega_c \gamma_{\rm neq}}
{\pi} [g'(0)]^2 \; \; .
\end{equation}

The appearance of the reactive frequency $\lambda$ is suggestive of the fact 
that the particle on the average is not moving on the bare barrier with
frequency $\omega_{b}$ but on a dressed barrier frequency $\bar{\omega}_{b}
$ corrected by $\lambda$. Pollak$^{8}$ has shown that the reactive frequency 
$\lambda$ is exactly an imaginary frequency of a barrier that has been modified
by the bath modeled as a discrete set of harmonic oscillators linearly
coupled to the system. The effect of $\lambda$ is to slow down the particle
by friction near the barrier. In the present model where the generalized
Langevin equation(12) describes the motion of the particle over a 
fluctuating barrier the essential modification of $\lambda$ and $\omega_{b}$
rests on the nonlinear coupling of the nonequilibrium relaxing modes with the
system. Thus in addition to the properties of the bath, dynamic nature of the
system-bath coupling is also significant in governing the barrier dynamics.
We note in passing that the usual Markovian limit can be recovered if one
puts $\gamma_{\rm neq}=0$ in Eq.(67) and associated quantities.

Before closing this section one pertinent point need to be mentioned. A closer
look into the derivation makes it clear that Eq.(67) results from an ansatz
of the form (53) where we use $\tilde{V}(x)$ in the Boltzmann factor. This
choice is basically guided by the fact that the potential $V(x)$ gets dressed
at $t=0$ by initial excitation of nonequilibrium modes. This choice also
 makes the stationary current independent of position. However, if one
uses the bare potential $V(x)$ and assume a weak dependence of $x$ on $j_{s}$,
one obtains Eq.(67) with $\bar{\omega}_{b}$ in the denominator getting
replaced by $\omega_{b}$ itself. The main lesson is that the modification of
Kramers rate (67) is essentially due to $\lambda$, the reactive frequency
of Grote-Hynes, which has been recognized as an important result in view of 
some experimental evidence$^{18}$ of relatively weak dependence of rate on 
damping in the large friction limit.

\vspace{0.5cm}

\begin{center}
\bf{V.\hspace{0.2cm}Time-dependent solution of the generalized Fokker-Planck
equation ; nonstationary Kramers rate ; nonexponential relaxation kinetics}
\end{center}

\vspace{0.5cm}

We now turn to Eq.(55). Rearranging the time-dependent $\Delta$-containing
terms it may be rewritten as 
\begin{equation} 
\frac{1}{\Gamma}\frac{\partial F}{\partial t}=-\left[\frac{ (\Gamma+a)v+\bar
{\omega}_{b}^{2}x}{\Gamma}\right]\frac{\partial F}{\partial u}+KT
\frac{\partial^{2} F}{\partial u^{2}}+\Delta\left[\frac{aKT}{\Gamma}
\frac{\partial^{2} F}{\partial u^{2}}-\frac{(av-\bar{\omega}_{b}^{2}x)}
{\Gamma}\frac{\partial F}{\partial u}\right]\hspace{0.2cm},
\end{equation} 

\noindent
where $\Delta$ is defined in Eq.(57).

Let us write
\begin{equation}
\frac{ (\Gamma+a)v+\bar{\omega}_{b}^{2}x}{\Gamma}=-\alpha u
\end{equation}
\noindent
and
\begin{equation}
\frac{(av-\bar{\omega}_{b}^{2}x)}{\Gamma}=-\lambda u
\end{equation}

\noindent
Here $\alpha$ is as defined in (63) and $\lambda$ is to be determined.

In terms of the relations (72) and (73), Eq.(71) reduces to a more 
compact form.
\begin{equation}
\frac{1}{\Gamma}\frac{\partial F}{\partial t}=\alpha u
\frac{\partial F}{\partial u}+KT\frac{\partial^{2} F}{\partial u^{2}}
+\Delta\left[\frac{aKT}{\Gamma}
\frac{\partial^{2} F}{\partial u^{2}}+\lambda u
\frac{\partial F}{\partial u}\right]\hspace{0.2cm}.
\end{equation}

Eq.(72) may be used to calculate the value of $a$ as obtained from the
solution of the algebraic equation
\begin{equation}
a^{2}+\Gamma a-\bar\omega_{b}^{2}=0\hspace{0.2cm}.
\end{equation}

Only the negative root of the above equation (say $a_{-}$) is the physically
realizable solution corresponding to the steady state solution. This value of
$a$ determines uniquely the value of $\lambda$ as defined in Eq.(73) to 
obtain
\begin{equation}
\lambda=-\alpha\hspace{0.2cm}.
\end{equation}

We now seek a solution $F(u,t)$ of Eq.(74) in the form
\begin{equation}
F(u,t)=F_{s}(u)e^{-\phi(t)}\hspace{0.2cm},
\end{equation}

\noindent
where $F_{s}(u)$ is the steady state solution obtained in the earlier section,
i.e., it satisfies
\begin{equation}
\alpha u \frac{\partial F_{s}}{\partial u}+KT\frac{\partial^{2}F_{s}}
{\partial u^{2}}=0\hspace{0.2cm}\hspace{0.2cm}.
\end{equation}

We require further
\begin{equation}
\left.{\cal L}t\right._{t\rightarrow\infty}\phi (t)=0\hspace{0.2cm}.
\end{equation}

\noindent
Substituting (77) in Eq.(74) it may be shown that the `space' and the time
part is separable. We obtain,
\begin{equation}
-\frac{1}{\Gamma}\frac{\partial \phi}{\partial t} e^{\frac{\gamma}{2} t}=\frac{C}
{F_{s}}\left[\lambda u\frac{\partial F_{s}}{\partial u}+\frac{aKT}{\Gamma}
\frac{\partial^{2}F_{s}}{\partial u^{2}} \right]={\rm constant}=D({\rm say})
\hspace{0.2cm},
\end{equation}
                                                                         
\noindent
where we have made use of the Eq.(78) and also 
\begin{eqnarray*}
\Delta=Ce^{-\frac{\gamma}{2} t}\hspace{0.2cm}
{\rm with} \hspace{0.2cm}C=\frac{r \gamma_{\rm neq}}{\gamma_{\rm eq}+\gamma_{
\rm neq}[g'(0)]^{2}}\hspace{0.2cm}.
\end{eqnarray*}

On integration over time we obtain from Eq.(80), the solution
\begin{equation}
\phi (t)=2 D\frac{\Gamma}{\gamma}e^{-\frac{\gamma}{2} t}
\end{equation}

\noindent
where $D$ is determined by the initial condition.

\noindent
The time-dependent solution of Eq.(71) therefore reads as
\begin{equation}
F(u,t)=F_{s}(u)\exp\left[-\frac{2D\Gamma}{\gamma}
e^{-\frac{\gamma}{2} t}\right]
\hspace{0.2cm}.
\end{equation}

\noindent
Thus the corresponding probability distribution is given by,
\begin{eqnarray}
p(x,v,t)=N\left[\left(\frac{\pi KT}{2\alpha}\right)^{\frac{1}{2}}+\int_{0}^
{v-|a|x}dz \exp\left(-\frac{\alpha z^{2}}{2KT}\right)\right]\nonumber\\
\nonumber\\
\exp\left[-\frac{\frac{v^{2}}{2}+\tilde{V}(x)}{KT}\right]
e^{-\frac{2D\Gamma}{\gamma}\left[\exp(-\frac{\gamma}{2} t)\right]}\hspace{0.2cm}.
\end{eqnarray}

To determine $D$ we now demand that just at the moment 
the system (and the nonthermal
bath) is subjected to external excitation at $t=0$ and $x\rightarrow
-\infty$ the distribution (75) must coincide with the usual Boltzmann 
distribution where the energy term in the Boltzmann factor in addition to
usual kinetic and potential terms contains the initial fluctuation of energy
density $\Delta {\cal U}$ 
[$\Delta {\cal U}={\cal U}(\omega,0)-\frac{1}{2} KT$] 
due to excitation of the system at $t=0$ [see Eq.(15)].

\newpage

\begin{eqnarray}
p(x,v,t)\stackrel{t\rightarrow 0}
{\longrightarrow} N\left(\frac{2\pi KT}{\alpha}\right)^{\frac{1}{2}}
e^{-2D\frac{\Gamma}{\gamma}}
e^{-\frac{1}{KT}\left(\frac{v^{2}}{2}+\tilde{V}(x)\right)}\nonumber\\
\nonumber\\
=N\left(\frac{2\pi KT}{\alpha}\right)^{\frac{1}{2}}
e^{-\frac{1}{KT}\left(\frac{v^{2}}{2}+{\tilde{V}}(x)+\Delta {\cal U}\right)}, 
\hspace{0.2cm}{\rm for} (x\rightarrow -\infty)\hspace{0.2cm}.
\end{eqnarray}

The last equality demands that
\begin{equation}
D=\frac{\gamma}{2\Gamma} \; \frac{\Delta {\cal U}}{KT}
\end{equation}

\noindent
[for the current to be coordinate independent the parabolic approximation
of $\tilde{V}(x)$ is to be used]. $D$ is thus determined in
terms of the relaxing mode parameters and fluctuations of the energy density 
distribution at $t=0$.

The time-dependent probability density therefore allows us to construct
nonstationary current,
\begin{equation}
j(t)=\int_{-\infty}^{+\infty}dv\hspace{0.1cm}v\hspace{0.1cm}p(x,v,t)=
j_{s}e^{-\frac{2D\Gamma}{\gamma}\exp(-\frac{\gamma}{2} t)}\hspace{0.2cm},
\end{equation}

\noindent
where $j_{s}$ is the stationary or steady state current as derived in the last
section.

By Eq.(74) we have,
\begin{equation}
p_{w}(x,v)=p(x\rightarrow -\infty,v,t=0_{-})\hspace{0.2cm},
\end{equation}

\noindent
which was used to calculate the number of particles $n_{a}$ initially in the 
well just before the system was subjected to shock at $t=0$. Thus 
non-stationary Kramers rate of transition is given by
\begin{equation}
k(t)=\frac{\omega_{0}}{2\pi\bar{\omega}_{b}}\left[\left\{\left(\frac{\Gamma}
{2}\right)^{2}+\bar{\omega}_{b}^{2}\right\}^{\frac{1}{2}}-\frac{\Gamma}{2}
\right]e^{-\frac{E_{b}}{KT}}
e^{-\left[\frac{2D\Gamma}{\gamma}\exp(-\frac{\gamma}{2} t)\right]}\hspace{0.2cm},
\end{equation}

\noindent
or in terms of the steady state Kramers rate $k$
\begin{equation}
k(t)=k\exp\left[ -\frac{\Delta {\cal U}}{KT} 
e^{-\frac{\gamma}{2} t}\right]\hspace{0.2cm} \; \; ,
\end{equation}

\noindent
where $\Delta {\cal U}$ is a measure of the initial departure from the average
energy density distribution due to the preparation of the nonstationary state of 
the intermediate bath modes as a result of excitation at $t=0$, and $k$ is
given by
\begin{equation}
k=\frac{\omega_0}{2\pi {\bar{\omega}}_b} \left[ \left \{ \left(
\frac{\Gamma}{2} \right )^2 + {\bar{\omega}}_b^2 \right \}^{1/2} -
\frac{\Gamma}{2} \right ] \;  e^{-E_b/KT} \; \; .
\end{equation}

The above result (88) illustrates a strong nonexponential relaxation of the system
mode undergoing a nonstationary activated rate process. The origin of this is an initial 
preparation of nonequilibrium mode density distribution (with a deviation 
$\Delta {\cal U}$) which eventually
relaxes to an equilibrium distribution. Eq. (88) implies that the initial 
transient rate is different from the asymptotic steady state Kramers rate. 
What is immediately apparent is that the sign of 
$\Delta {\cal U} [ = {\cal U}(\omega,0)-\frac{1}{2}KT ]$ 
determines whether the initial
rate will be faster or slower than the steady state rate. When 
$\Delta {\cal U}$
is negative, i.e., the contribution of thermal energy dominates, the initial 
rate of thermal activation of the reaction co-ordinate gets enhanced as 
a consequence.
On the other hand, when the sudden excitation of the nonequilibrium
modes provides a positive deviation $\Delta {\cal U}$, the initial rate of 
activation becomes slower. This is because there likely to exist some
time lag for the nonthermal energy gained by the few nonequilibrium modes 
by sudden 
excitation to be distributed over a range before it become available 
to the reaction co-ordinate as thermal energy for activation.

It is also interesting to consider the zero and high temperature limits. 
When $T\rightarrow 0$ both the steady state Kramers rate $k$ as well
as the time-dependent factor 
$exp[-\frac{\Delta {\cal U}}{KT} e^{-\gamma t/2} ]$
goes to zero. If $T=0$, then $k(t)$ is zero at all time. However, it seems
intuitively that there should be a transient period during which the rate is 
finite. It may be noted that since the relaxation of the nonequilibrium 
bath modes (following the sudden excitation) is very slow compared to the 
rate of activation process, the particle undergoing barrier crossing cannot
`sense' this transient (ideally if the relaxation to equilibrium is adiabatic,
i.e., the thermalization of the initial departure $\Delta{\cal U}$ is very 
slow, there should be no transient). We believe that the distinct separation
of the two time-scales implied in the dynamics makes the transient
unobservable. An interplay of overlapping time-scales pertaining to 
the relaxation of
the bath and the activation of the system may give rise to transients in 
$k(t)$ at $T=0$. Evidently this is outside the scope of the present treatment.

When $T$ is very high such that $\frac{1}{2} KT$ far exceeds
${\cal U}(\omega,0)$ the initial rate gets strongly enhanced 
(since $\Delta {\cal U}$ is
negative) and the time-dependent exponential factor becomes roughly 
independent of temperature.
In the limit $t\rightarrow \infty$ or $\Delta {\cal U}\rightarrow 0$ we recover
steady state Kramers rate, as expected.
 
The activation rate is thus consequently modified which effectively 
incorporates a secondary relaxation kinetics. The quasi-thermal excitations
decay on the time scale $\frac{1}{\gamma}$, which is well separated from other
internal time scales of the thermal bath. The dynamic nature of the coupling
between the system and the nonequilibrium modes is responsible for fluctuating
barrier. 
A closer look into the origin of the non-exponential kinetics makes it clear
that the spiritual root of $D$-term is essentially the $\Delta$-containing
term in Eq.(71) or $\frac{\partial^{2} p}{\partial x\partial v}$ term in
Eq.(46) which is a non-Markovian contribution. We thus identify the
non-exponential relaxation of the system mode as a typical non-Markovian
dynamical feature. In the case of very small $\gamma$ one naturally recovers
the exponential relaxation and Arrhenious rate of activation of the usual
kinetic scheme.

A relevant pertinent point regarding some of the related works need be
considered here. Generalized Langevin equation (GLE) has been widely employed
in various contexts, e.g., in the description of reactions in liquids. A
search for realistic models began with the realization that friction exerted
by the solvent on the solute is space dependent. A formally consistent
approach to the problem of space and time dependent friction had been 
introduced early by Lindenberg and co-workers$^{20,21}$. Carmeli and 
Nitzan$^{22}$ have also derived a stochastic dynamical equation which is a 
generalization of GLE to the case of space and time dependent friction. 
Pollak and Berezhkorskii$^{25}$ have demonstrated that the space and
time-dependent friction model is identical to a multidimensional 
anisotropic but Markovian friction problem in which the reaction co-ordinate
is coupled to an additional co-ordinate which is governed by a Langevin type 
equation. A theory for treating spatially dependent friction in the classical
activated rate processes has been considered and following the method of
Pollak an effective Grote-Hynes reactive frequency for this case has been 
obtained as a transcendental equation$^{23}$. More recently a general theory
for thermally activated rate constants influenced by spatially dependent and
time correlated friction$^{24}$ has been proposed.

While in the above problems one is concerned with the space and time dependent
friction, which is essentially a characteristic of the solvent mode structure,
in the present problem we deal with effect of a secondary relaxation of
intermediate oscillator modes (following an initial excitation) on the
primary kinetics of the system mode. The mode density function due to initial
excitation differs from its equilibrium value - a feature which is marked in
the nonequilibrium fluctuation-dissipation relation. Thus the exponential relaxation
in Eq.(81) is not be confused with the exponential time-dependent friction
employed in earlier instances. The origin of these two exponential terms are
fundamentally different. The non-exponential kinetics is essentially an
offshoot of a dynamic modification of the fluctuation-dissipation theorem
appropriately carried over to a nonstationary regime. This nonequilibrium 
nature of activated process is reflected in the nonstationary kinetics that
we derive here.

The non - exponential relaxation kinetics had been 
explored earlier in different
occasions in relation  to  disordered  
systems$^{13}$,  viscous  liquids$^{19}$,  oxygen  binding  to 
h$\mbox{\ae}$moglobin$^{16}$, where
phenomenological fluctuating barrier models have been employed (barriers
arising from the collective motions of many degrees of freedom). The 
present model although oversimplified in many respects captures the essential
nature of influence of an initial non-thermal mode density distribution on the
relaxation kinetics of the system.

\begin{center}
\bf{VI.\hspace{0.2cm}Conclusions}
\end{center}

In conclusion, we consider a simple microscopic system-nonequilibrium
bath model to simulate nonstationary thermally activated processes.
The nonequilibrium bath is effectively realized in terms of a 
semi-infinite dimensional broad-band reservoir which is subsequently
kept in contact with a thermal reservoir which allows the nonthermal bath 
to relax with a characteristic time. A systematic separation of timescales
is then used to construct the appropriate Langevin equation for the
particle, which is nonlinear and non-Markovian in character.
Based on a strategy of Van Kampen's expansion in $\epsilon \tau_{c}$
of the relevant physical quantity where $\epsilon$ is the strength and
$\tau_{c}$ is the correlation time of fluctuations of the relaxing modes,
we show that this Langevin equation can be recast into the form of a
generalized Fokker-Planck equation, when the correlation time is short but 
finite. Adelman's form of the Fokker-Planck equation [ Eq.(8) ] as well as
the standard Markovian description can be recovered in the appropriate limits. We
now summarize the main conclusions of this study:

(i) The model proposed here captures the essential features of
Langevin dynamics with a fluctuating barrier. The present approach
is equipped to deal with situations both in the non-stationary short
time as well as stationary long time regimes. The origin of the short
time non-exponential kinetics can be traced back in a non-stationary 
fluctuation-dissipation theorem.

(ii) We derive the expression for the steady state Kramers escape rate 
in the non-Markovian case and show that the Grote-Hynes `reactive frequency'
can be realized explicitly in terms of the microscopic parameters of the  
nonequilibrium relaxing modes and their arbitrary dynamic coupling to the 
system mode.

(iii) The central result of this paper is the derivation of a nonstationary
Kramers rate in closed analytic form. This essentially illustrates the
influence of an initial excitation and subsequent relaxation of the
nonequilibrium bath modes on the system degree of freedom undergoing
an activated process. The system mode is shown to follow strong 
non-exponential kinetics.

The model considered in the present paper may be realized in a guest-host
system embedded in a lattice where the immediate local neighborhood of the 
guest comprises intermediate oscillator modes whereas the lattice plays
the role of a thermal bath. Appropriately identified reaction co-ordinate 
coupled to other degrees of freedom in a molecule
embedded in a matrix may be another worthwhile candidate for such a scheme.

Although simple, the model thus allows us explicit solutions and in  
view of the prototypical role played by the present model in several earlier 
investigations, we hope that the conclusions drawn here will find 
applications in some related experiments of physics and chemistry of complex
systems.

{\bf{Acknowledgments}} : Partial financial support from the Department of
Science and Technology (Govt. of India) is thankfully acknowledged. One of the
authors (JRC) is thankful to Prof. J. K. Bhattacharjee (Dept. of Theoretical
Physics, I.A.C.S) and to S. K. Banik for helpful discussions and suggestions.

\newpage

\begin{center}
{\large {\bf References}}
\end{center}

\vspace{0.5cm}

\begin{enumerate}
\item H. A. Kramers, Physica {\bf 7}, 284 (1940).
\item See for example, a review by P. H\"anggi, P. Talkner and M. Borkovec,
Rev. Mod. Phys. {\bf 62}, 251 (1990) and the references given therein.
\item J. T. Hynes and J. M. Dentch, {\it Physical Chemistry : An Advanced
Treatise}, vol. IIB (Academic, New York, 1975) ; S. A. Adelman and J. D. Doll,
Acc. Chem. Res. {\bf 10}, 378 (1997) ; J. M. Dentch and I. Oppenheim, 
J. Chem. Phys. {\bf 54}, 3547 (1971).
\item R. F. Grote and J. T. Hynes, J. Chem. Phys. {\bf 73}, 2715 (1980).
\item S. A. Adelman, J. Chem. Phys. {\bf 64}, 124 (1976).
\item P. H\"anggi and F. Mojtabai, Phys. Rev. A {\bf 26}, 1168 (1982) ;
P. H\"anggi, J. Stat. Phys. {\bf 30}, 401 (1983).
\item B. Carmeli and A. Nitzan, J. Chem. Phys. {\bf 79}, 393 (1983) ; Phys.
Rev. Letts. {\bf 49}, 423 (1982) ; Phys. Rev. A {\bf 29}, 1481 (1984).
\item E. Pollak, J. Chem. Phys. {\bf 85}, 865 (1986).
\item W. H. Louisell, {\it Quantum Statistical Properties of Radiation}
(Wiley, New York, 1973) ; M. Lax, Phys. Rev. {\bf 145}, 110 (1966) ;
R. Graham and H. Haken, Z. Phys. {\bf 235}, 166 (1970).
\item M. Millonas and C. Ray, Phys. Rev. Letts. {\bf 75}, 1110 (1995).
\item See, for example, a review by G. Gangopadhyay and D. S. Ray,
{\it Advances in multiphoton processes and spectroscopy} (vol. 8), ed. by
S. H. Lin, A. A. Villayes and F. Fujimura (World Scientific, Singapore, 1993)
; G. Gangopadhyay and D. S. Ray, Phys. Rev. A {\bf 46}, 1507 (1992) ; Phys.
Rev. A {\bf 43}, 6424 (1991) ; J. Chem. Phys. {\bf 96}, 4693 (1992).
\item N. G. van Kampen, Phys. Reps. {\bf 24}, 171 (1976).
\item R. Landauer, J. App. Phys. {\bf 33}, 2207 (1962) ; J. Stat. Phys.
{\bf 9}, 351 (1973).
\item D. L. Stein, C. R. Doering, R. G. Palmer, J. L. van Hemmen and R. M.
McLaughlin, J. Phys. A {\bf 23}, L203 (1990).
\item P. Reimann and T. C. Elston, Phys. Rev. Letts. {\bf 77}, 5328 (1996).
\item M. Karplus and J. A. McCamaron, CRC Crit. Rev. Biochem. {\bf 9}, 293
(1981).
\item J. Wong and C. A. Angell, {\it Glass : Structure by Spectroscopy}
(New York, Dekker, 1976).
\item S. P. Velsko, D. H. Waldeck and G. R. Fleming, J. Chem. Phys. {\bf 78},
294 (1983).
\item C. A. Angell, {\it Relaxation in Complex System}, ed. by K. L. Ngai and
G. B. Wright (Washington DC. US Govt. printing office, 1985).
\item K. Lindenberg and V. Seshadri, Physica {\bf 109A}, 483 (1981).
\item K. Lindenberg and E. Cortes, Physica {\bf 126A}, 489 (1984).
\item B. Carmeli and A. Nitzan, Chem. Phys. Letts. {\bf 102}, 517 (1983).
\item G. A. Voth, J. Chem. Phys. {\bf 97}, 5908 (1992).
\item G. R. Hynes, G. A. Voth and E. Pollak, J. Chem. Phys. {\bf 101},
7811 (1994).
\item E. Pollak and A. Berezhkorskii, J. Chem. Phys. {\bf 99}, 1344 (1993).
\end{enumerate}

\end{document}